\documentstyle[twoside,fleqn,espcrc2,psfig]{article}

\newcommand{\AmS}{{\protect\the\textfont2
  A\kern-.1667em\lower.5ex\hbox{M}\kern-.125emS}}

\title{Forces between composite particles in QCD}

\author{H.R. Fiebig\address{Physics Department, 
                            FIU - University Park, 
                            Miami, Florida 33199, USA},
        H. Markum\address{Institut f\"ur Kernphysik,
                            Technische Universit\"at Wien,
                            A-1040 Vienna, Austria},
        A. Mih\'aly\address{Department of 
                            Theoretical Physics, 
                            Lajos Kossuth University,
                            H-4010 Debrecen, Hungary}
        and 
        K. Rabitsch$^{\mbox{\scriptsize b}}$
                                 }
       
\begin{document}

\begin{abstract}
Starting from the meson-meson Green function in 3+1 dimensional 
quenched lattice QCD we calculate potentials between heavy-light mesons
for various light-quark mass parameters. 
For the valence quarks we employ the staggered scheme. 
The resulting potentials turn out to be short ranged and attractive. 
A comparison with a tadpole improved action for the gauge fields is presented.
\end{abstract}

\maketitle

\section{INTRODUCTION}

    For several decades nucleon-nucleon interactions have been parametrized by
phenomenological potentials. Substantial effort, employing purely hadronic
degrees of freedom, has lead to meson-exchange potentials \cite{machleidt}. 
Attempts to take into account quark and gluon degrees of freedom with hybrid 
models have also been made \cite{faessler}.

    Today, QCD is believed to be the fundamental theory of strong interactions.
Thus it has become a challenge for theoretical nuclear physics to extract a
nucleon-nucleon potential from first principles. In the low energy regime of QCD
nonperturbative tools have to be used. This leads us to study the forces in
systems of two hadrons on the lattice.

    Previous lattice calculations with static valence quarks have revealed an
attractive force between two three-quark clusters \cite{rabitsch}. 
When dynamical quark propagators are used antisymmetrization and the 
exchange of valence quarks become possible \cite{Can94}. 
In 2+1 dimensional QED the potential between light mesons
exhibits a repulsive hard core and is attractive at intermediate distances 
\cite{Fie96}.  An extension of this formalism to 3+1 dimensional QCD 
using a hopping parameter expansion for the quark propagators is 
reported in \cite{Fie96a}.

    In the present work we take another step toward the goal of calculating
hadronic potentials from QCD. We study a system of two heavy-light
mesons with dynamical quark propagators for the light valence quarks.
Results from calculations with the Wilson action for the gauge field and
a tadpole improved gauge field action are reported.

\section{THEORY}

\subsection{Meson-meson correlator}

The one-meson field is a product of staggered Grassmann fields 
$\chi$ and $\bar{\chi}$ with a heavy and a light external flavor
$h$ and $l$, respectively,
\begin{equation}
\phi_{\vec{x}}(t) = \bar{\chi}_l(\vec{x}t) \chi_h(\vec{x}t) \,.
\label{eq1}\end{equation}
The meson-meson fields with relative distance $\vec{r}=\vec{y}-\vec{x}$ 
are then constructed by
\begin{equation}
\Phi_{\vec{r}}(t) = V^{-1} \sum_{\vec{x}} \sum_{\vec{y}} 
\delta_{\vec{r},\vec{y}-\vec{x}} \phi_{\vec{x}}(t) \phi_{\vec{y}}(t) \,. 
\label{eq2}\end{equation}
 
The information about the dynamics of the meson-meson system is
contained in the time correlation matrix
\begin{equation}
C_{\vec{r}\vec{r}\,'}(t,t_0) = 
\langle \Phi^{\dagger}_{\vec{r}}(t) \Phi_{\vec{r}\,'}(t_0) \rangle 
 - \langle \Phi^{\dagger}_{\vec{r}}(t) \rangle 
   \langle \Phi_{\vec{r}\,'}(t_0) \rangle , 
\label{eq3}\end{equation}
where $\langle \; \rangle$ denotes the gauge field configuration average. 
On the hadronic level $C$ is a 2-point correlator of a 
composite local operator describing a molecule-like structure.
Working out the contractions between the Grassmann fields the following
diagrammatic representation is obtained
\begin{equation}
C       = C^{(A)} - C^{(C)}  
        =
\mbox{ \begin{picture}(18,17)
       \linethickness{0.4mm}
       \put(2,-6){\line(0,1){15}}
       \thinlines
       \put(4,-6){\line(0,1){15}}
       \put(11,-6){\line(0,1){15}}
       \linethickness{0.4mm}
       \put(13,-6){\line(0,1){15}}
       \thinlines
       \put(2,-6){\line(1,0){2}}
       \put(2,9){\line(1,0){2}}
       \put(11,-6){\line(1,0){2}}
       \put(11,9){\line(1,0){2}}
       \end{picture} }
-
\mbox{ \begin{picture}(18,17)
       \linethickness{0.4mm}
       \put(2,-6){\line(0,1){15}}
       \thinlines
       \put(3,-6){\line(3,5){9}}
       \put(12,-6){\line(-3,5){9}}
       \linethickness{0.4mm}
       \put(13,-6){\line(0,1){15}}
       \put(2,4){\vector(0,1){0}}
       \thinlines
       \put(2,-6){\line(1,0){1}}
       \put(2,9){\line(1,0){1}}
       \put(12,-6){\line(1,0){1}}
       \put(12,9){\line(1,0){1}}
       \end{picture} } \, .
\label{eq5}
\end{equation}
Each contribution to the correlator comprises the exchange 
of gluons. Thus, even diagram $C^{(A)}$ leads to forces,
whereas diagram $C^{(C)}$  
corresponds also to interaction mediated by the exchange of 
the light valence quark.
Denoting contractions of the Grassmann fields by
\begin{equation}
\ldots \stackrel{n}{\chi} \ldots \stackrel{n}{\bar{\chi}} \ldots\: = \: 
G_n \,, \quad\mbox{with}\quad n=1 \ldots 4 \,, 
\label{eq6}\end{equation}
we have for example
\begin{equation}
C^{(A)} \sim \langle \stackrel{43\phantom{99}}{\phi^{\dagger}_{\vec{y}\,'}}
                 \stackrel{21\phantom{99}}{\phi^{\dagger}_{\vec{x}\,'}}
                 \stackrel{12\phantom{9}}{\phi^{\phantom{\dagger}}_{\vec{x}}}
                 \stackrel{34\phantom{9}}{\phi^{\phantom{\dagger}}_{\vec{y}}}
              \rangle
= \langle G_1^{(h)} G^{\ast (l)}_2 G_3^{(h)} G^{\ast (l)}_4 \rangle ,
\label{eq78}\end{equation}
with $\vec{r}=\vec{y}-\vec{x}$, $\vec{r}\,'=\vec{y}\,'-\vec{x}\,'$,
and $\sim$ stands for the sums and factors that carry over 
from (\ref{eq2}). Color indices are suppressed.
The gauge configuration average is taken over the
product of all four propagators $G$. 
The propagator of the light quark $G^{(l)}$ 
is obtained from inverting the staggered
fermion matrix with a random source estimator.
A standard conjugate gradient algorithm is used.
The heavy-quark propagator is given by
\begin{equation}
G^{(h)}_{\vec xt,\vec xt_0} = \left(\frac{1}{2m_{h}a}\right)^k
[\Gamma_{\vec x 4}]^k
\prod_{j=1}^kU_{x=(\vec x,ja),\mu =4} \, ,
\label{heavyqprop}
\end{equation}
where the phase factors $\Gamma_{\vec x 4}=(-1)^{(x_1+x_2+x_3)/a}$
in the Kogut-Susskind formulation
correspond to the Dirac matrices and $k=(t-t_0)/a$.
In our calculations we set $m_{h}a=1$.

Since the heavy valence quarks are fixed in space
the relative distance between the mesons 
is the same at the initial and final time of the propagation, 
$\vec{r}\,'=\vec{r}$.
The effective ground-state energy of the meson-meson
system $W(\vec r)$ can then be extracted from the large euclidean
time behavior of $C$ following quantum-mechanical
reasoning for composite particles \cite{Fie96a},
\begin{eqnarray}
C_{\vec{r}\vec{r}}(t,t_0) 
& = & \sum_{n} \left| \langle \vec{r} | n \rangle \right| {}^2 
           e^{-E_n (t-t_0)} \nonumber\\
& \simeq & c(\vec{r}\,) e^{-W(\vec{r})(t-t_0)} \,.
\label{eq15}
\end{eqnarray}
The residual meson-meson potential is
\begin{equation}
V(r)=W(r)-2m \, ,
\end{equation}
with the mass $2m$ of two free mesons subtracted.

\subsection{Improved action}

Discretization errors due to finite lattice spacing $a$ can be reduced
by including terms of higher order in $a$ with the action.
It has been shown that at the classical level adding a term with
six-link rectangular plaquettes $U_{rt}$ to the usual Wilson
gauge field action removes $O(a^{2})$ errors \cite{weisz}.
A significant improvement is obtained by introducing tadpole factors
in the six-link term \cite{alford}. We use the improved action
\begin{eqnarray}
S[U]  & = &  \beta_{pl}\sum_{pl}\frac13 \mbox{\rm Re}\,\mbox{\rm Tr}(1-U_{pl})
\nonumber \\
& + & \beta_{rt}\sum_{rt}\frac13 \mbox{\rm Re}\,\mbox{\rm Tr}(1-U_{rt})\, ,
\end{eqnarray}
where the first term is the Wilson action with four-link
plaquettes $U_{pl}$. The coupling parameter
\begin{equation}
\beta_{rt} = - \frac{\beta_{pl}}{20 u_0^2} 
\end{equation}
and the mean link
\begin{equation}
u_0 = (\frac{1}{3} \mbox{\rm Re}\,\mbox{\rm Tr} \langle U_{pl}\rangle)^{1/4}
\end{equation}
are determined self-consistently for a given $\beta_{pl}$.

\begin{figure*}[tp]
\refstepcounter{figure}
\label{figconvimp}
\hbox{{\bf \hspace*{29mm} Wilson \hspace{57mm} Tadpole improved}}
\vspace{2mm}
\centerline{\hbox{
\psfig{figure=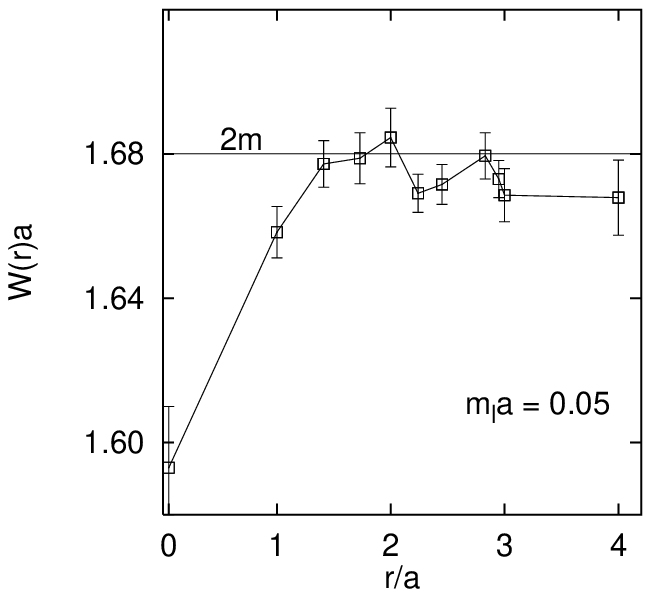,width=65mm,height=59mm,rwidth=80mm}
\psfig{figure=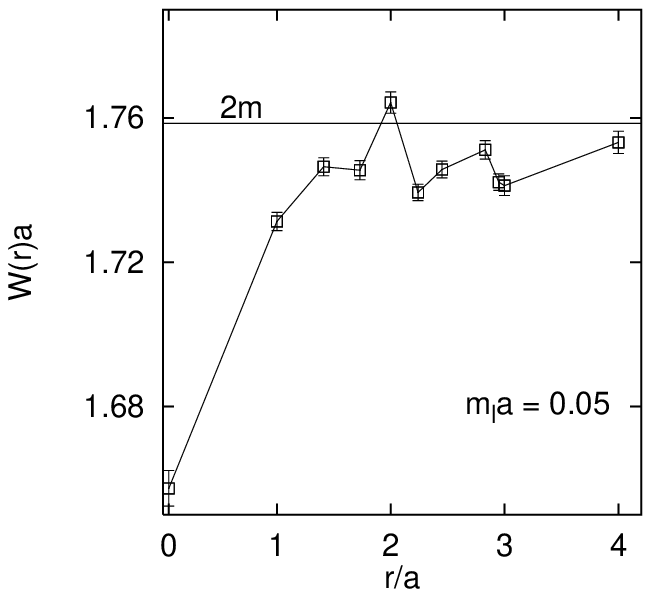,width=65mm,height=59mm,rwidth=80mm}
}}
\centerline{\hbox{
\psfig{figure=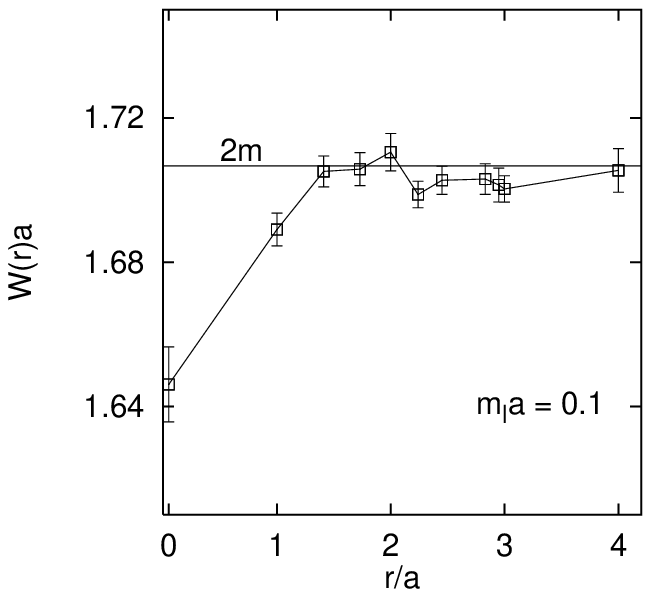,width=65mm,height=59mm,rwidth=80mm}
\psfig{figure=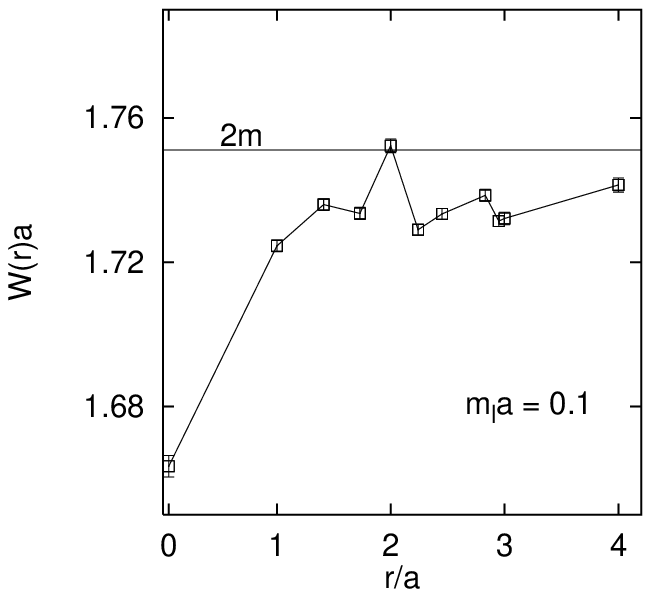,width=65mm,height=59mm,rwidth=80mm}
}}
\centerline{\hbox{
\psfig{figure=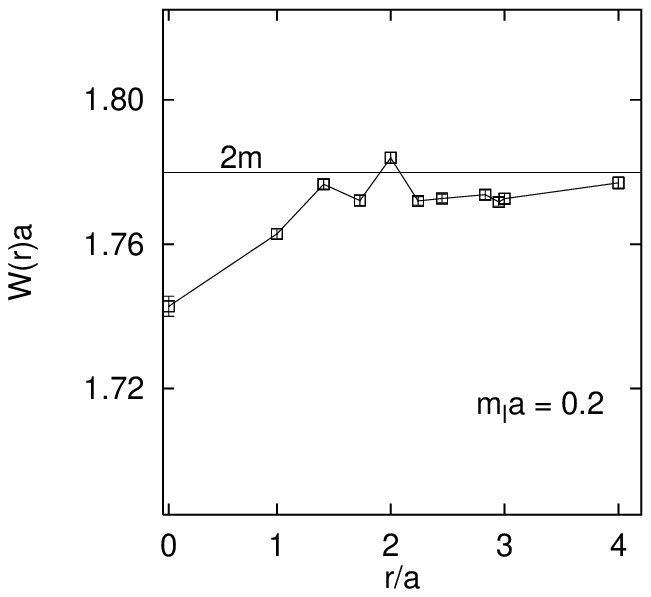,width=65mm,height=59mm,rwidth=80mm}
\psfig{figure=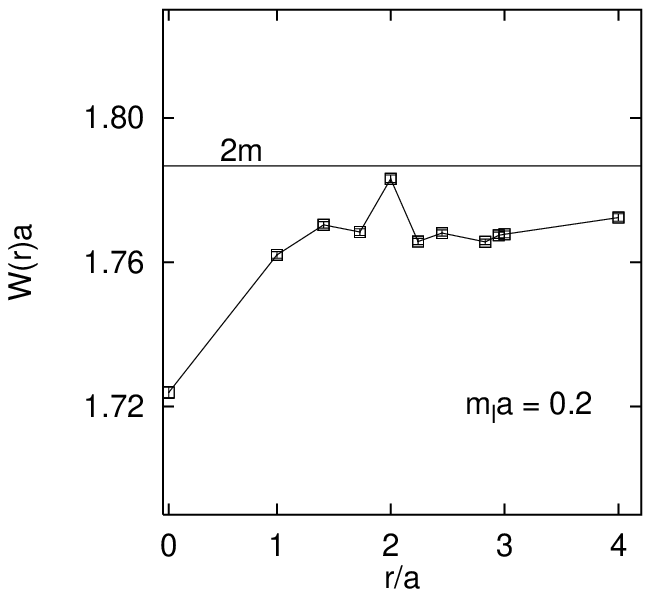,width=65mm,height=59mm,rwidth=80mm}
}}
{\noindent
Figure~\ref{figconvimp}: 
Potentials between heavy-light mesons obtained from
simulations with conventional Wilson action 
compared to results with improved action. 
The potentials are attractive and become stronger with decreasing 
light-quark mass. Error bars are smaller when using tadpole
improved action. Improvement was restricted to the gauge field action.
}
\end{figure*}

\section{RESULTS}

We considered a periodic $8^3\times16$ lattice. Each potential is the
result of a measurement on 100 independent gauge field configurations
which were separated by 200 sweeps. 
The inversion of the fermion matrix was performed with 32 random sources.

The effective energy $W(r)$ was extracted from the correlator $C$
by a four parameter (Levenberg-Marquardt) fit with the function
\begin{eqnarray}
C_{rr}(t) & = & B(r)\cosh{[ W(r)(t-8a)]}  \nonumber\\
 & + & (-1)^{t/a}
\widetilde{B}(r)\cosh{[ \widetilde{W}(r)(t-8a)]} .
\end{eqnarray}
The second term alternating in sign is a peculiarity of the
staggered scheme.
In order to improve the quality of the fits 15 time slices were used.
The extracted potentials are thus between meson states which are contaminated
by excitations.
To be numerically consistent with $W(r)$ at large distances, the mass $2m$ 
of two noninteracting mesons was extracted from the square of
the meson two-point function $[C^{(2)}]^{2}$.

\subsection{Wilson action}

For the update with the usual Wilson action we chose the
inverse gluon coupling constant $\beta = 5.6$.
This corresponds to a lattice spacing of approximately 0.2fm.
The left column of Fig.~\ref{figconvimp} shows the resulting potentials
for different values of the light valence quark mass 
$m_{l}a=0.05, 0.1$ and $0.2$, respectively.
In all cases the potential between two
heavy-light mesons is short ranged and attractive.
Calculations from inverse scattering theory propose a similar shape
for $\mbox{K}\bar{\mbox{K}}$-potentials \cite{sander96}.
Experimental scattering data contain inelastic channels 
which are not accounted for in our QCD calculation.
At distances $r/a>1$ essentially no interaction energy 
can be resolved and the effective
total energy approaches the value of two noninteracting mesons $2m$.
With decreasing light-quark mass the interaction becomes stronger.

\subsection{Improved action}

Here the coupling was set to $\beta_{pl}=7.6$.
This value corresponds to $a \approx 0.2$fm, as for the
Wilson action \cite{alford}.
The right column of
Fig.~\ref{figconvimp} shows the resulting potentials, again
for different values of the light valence quark mass
$m_{l}a=0.05, 0.1$ and $0.2$. 
Again we obtain attractive potentials which become deeper with decreasing
quark masses. Those exhibit larger anisotropies than in the
Wilson case. The likely reason is that only the gluonic action but not
the staggered fermion matrix has been improved. 
In any case, it is interesting to observe that the shape of the
potential remains stable.

\section{CONCLUSIONS}

We applied a practical method to extract effective hadron-hadron potentials
from lattice QCD to a system of two heavy-light mesons.
For the light valence quark we employed dynamical quark propagators.
The resulting potentials are short ranged and attractive.
Smaller quark mass parameters lead to stronger interaction.
A preliminary analysis of simulations with improved action showed 
consistent results.
Further insight into the interaction mechanism is expected from
calculations with mesons consisting of two light valence quarks.

\section{ACKNOWLEDGEMENTS}

This work was supported in part by the Research Group in Physics of the 
Hungarian Academy of Sciences, Debrecen,
by ``Fonds zur F\"orderung der wissenschaftlichen Forschung'' 
under Contract No.~P10468-PHY and
by NSF grant PHY-9409195.
The hospitality of CEBAF/TJNAF is greatly acknowledged.


\begin{thebibliography}{9}
\bibitem{machleidt} R.~Machleidt, Adv.~Nucl.~Phys. 19 (1989) 189.
\bibitem{faessler} A.~Faessler, F.~Fernandez, G.~L\"ubeck and K.~Shimizu, 
                Nucl.~Phys. A402 (1983) 555; 
                C.E.~DeTar, Lect.~Notes~in~Phys. 87, Springer (1978) 113. 
\bibitem{rabitsch}
                K.~Rabitsch, H.~Markum and W.~Sakuler, Phys.~Lett. B318 
                (1993) 507. 
\bibitem{Can94} J.D.~Canosa and H.R.~Fiebig, Nucl.~Phys. B~(Proc.~Suppl.)~34 
                (1994) 561.
\bibitem{Fie96} H.R.~Fiebig, O.~Linsuain, H.~Markum and K.~Rabitsch, 
                Nucl.~Phys. B~(Proc.~Suppl.)~47 (1996) 695.
\bibitem{Fie96a}H.R.~Fiebig, H.~Markum, A.~Mih\'aly, K.~Rabitsch, 
                W.~Sakuler and C.~Starkjohann,
                Nucl.~Phys. B~(Proc.~Suppl.)~47 (1996) 394.
\bibitem{weisz} P.~Weisz and R.~Wohlert, Nucl.~Phys.~B236 (1984) 397.
\bibitem{alford} M.~Alford, W.~Dimm, G.P.~Lepage, G.~Hockney and 
                P.B.~Mackenzie,
                Nucl.~Phys. B~(Proc.~Suppl.)~42 (1995) 787;
                Phys.~Lett. B361 (1995) 87.
\bibitem{sander96} M.~Sander and H.V.~von~Geramb, 
                Nuclear Theory Report KTH--96/03, Univ.~of Hamburg  (1996).
\end{thebibliography}
\end{document}